\documentclass[12pt,a4paper]{article}   
\usepackage[dvips]{graphicx}
\usepackage{amsmath,amsfonts,amssymb,amsthm}  

\def\m@th{\mathsurround=0pt}
\def\pn{\par\noindent}
\newcommand{\np}{\nonumber\\}
\newtheorem{thm}{Theorem}
\newtheorem{lem}{Lemma}

\newtheorem{df}{Definition}
\newtheorem{cj}{Conjecture}

\title{
Stokes Multipliers, Spectral Determinants and T-Q relations 
  \footnote{based on talks given at
    APCTP conference, " Exactly Solvable Models of Statistical Mechanics
	and Mathematical Physics " (Seoul June 2000) and RIMS conference,
	"  Development in Discrete Integrable Systems - 
	Ultra-Discretization, Quantization  "
    (Kyoto August 2000)
           }
}
\author{ J. Suzuki\thanks{e-mail: sjsuzuk@ipc.shizuoka.ac.jp}\\
        \parbox{0.9\textwidth}{
        {\em
        \begin{center}
       Department of Physics, Faculty of Science\\
       Shizuoka University,\\
      Ohya 836, Shizuoka,\\
       Japan
        \end{center}
        }}
       }
\date{August 2000}
\begin{document}
\maketitle
\begin{abstract}
Recently, a remarkable correspondence
has been unveiled between a certain class of 
ordinary linear differential equations (ODE)
and integrable models.
In the first part of the report,
we survey the results concerning the 2nd order
differential equations,
the Schr{\"o}dinger equation
with a polynomial potential.
We will observe that 
fundamental objects in the study of the solvable models,
e.g., Baxter's $Q-$ operator, fusion transfer matrices 
come into play in the analyses on ODE.
The second  part of the talk is devoted to
the generalization to higher order linear differential equations. 
The correspondence found in the case of the 2nd order ODE is 
naturally lifted up.
We also mention a connection to the discrete soliton
theory.

\end{abstract}
\clearpage
%+++++++++++++++++++++++++++++++++++++++++++++++++++++++ Sec1

\section{Introduction}\label{intro}

Recent studies\cite{DT1}-\cite{doublewell} reveal an unexpected 
connection between  a certain category of ordinary
differential equations (ODE) and integrable models (IM) with quantum group symmetry.
We call this ODE/IM  correspondence \footnote{Following R. Tateo.}.
The success inherits the fruitful results from the exact 
 WKB analysis\cite{V1}-\cite{Takei} and progress in
 the study of integrable structure \cite{Baxbook}-\cite{BLZ3}.

The aim of the present talk is two-fold,
the survey on the 2nd order ODE case, 
and the brief preview on the generalization to 
higher order cases.

Firstly,  we will review  the results on the 2nd order
differential equations,
the Schr{\"o}dinger equation
with a polynomial potential,
$$
-\frac{d^2 y}{dx^2} + x^{\ell} y = E y.
$$
We will mainly follow the argument in \cite{DT2} but employ
some simplifications and add some materials.
Motivated by the success of the exact WKB method, we regard the
coordinate $x$ as a complex variable.
The complex $x$ plane is conveniently divided into
sectors. See section \ref{defStokes}.
Each sector possesses two 
(=  the order of the equation ) linear independent solutions.
We call them
a fundamental set of solutions (FSS).
The relations among FSS of different sectors are 
of our interest.
To be precise, we would like to evaluate the 
Stokes multiplier which characterizes the connection rule.
In this view point, it is natural to regard that the problem consists of
two coupled equations, the original differential equation
and the difference equation for  the Stokes multiplier.
It will be then shown that 
fundamental objects in the study of the solvable models,
e.g., Baxter's $Q-$ operator,
 fusion hierarchy of transfer matrices based on $U_q(A^{(1)}_1)$
and their functional relations
naturally come into play in the analyses on ODE.
Especially, (unfused) transfer matrix is identified with
the Stokes multiplier.
Reflecting the ODE/IM correspondence, the Stokes multiplier has 
two representations,
the Wronskian representation, which 
arises from ODE,
and the DVF representation, originated from IM.
They both play a role in generalizing the results 
in the second part of the talk.

Physically, the Stokes multiplier may be
less interesting.
Rather,  the quantity of importance is the spectral determinant,
$D(E)={\rm det}({\cal H}-E)$ or eigenvalues themselves.
Remarkably, $D(E)$ also belongs to 
the  fusion hierarchy.
Thus the result 
provides a unified view of Stokes multipliers and spectral determinant.\pn

In the second part, a generalization 
to higher order ODE will be addressed,

$$
-\frac{d^{n+1} y}{dx^{n+1}} + x^{\ell} y = E y.
$$

In \cite{SUNStokes}, functional relations were derived among 
Stokes multipliers and their generalizations.
These are identical to functional relations among
transfer matrices of solvable models with $U_q(A^{(1)}_n)$ symmetry,
which generalizes the observation for $n=1$.
The relations were evolved by use of the machinery in the
solvable models, a quantum analogue of the Jacobi-Trudi formula.
Here we will give an alternative, much simpler derivation, 
resulting from the Wronskian representation of the
Stokes multipliers.
We also note that the possible connection of the relations to
discrete soliton equations (the Hirota-Miwa equation)\cite{Hiro,Miwa}.

The parallelism to IM will be further
exploited.
The eigenvalues of transfer matrices possess a universal structure
called the dressed vacuum form (DVF).
The universality has a deep origin in analyticity of their expressions
under Bethe ansatz
equations and the Yang-Baxter integrability.
We will show that Stokes multipliers also assume the same DVF.

The paper is organized as follows.
The asymptotic form for $n+1-$th ODE will be
discussed in section \ref{defStokes}.
Several notations and symbols, such as sectors, Stokes matrices,
are introduced for $n$ general.
In the next three sections, we restrict ourselves to the $n=1$ case.
The recursion relations and functional relations
 for the Stokes multiplier and 
its generalizations are derived in section \ref{FRsl2}.
Under certain assumptions, one transforms the 
algebraic relations to a set of integral equations modulo one unknown 
parameter.
Remarkably, the integral equations take identical forms to 
thermodynamic Bethe ansatz equations.
We shall discuss the DVF representation for the Stokes multiplier
in section \ref{dvf}.
The spectral problem is addressed in section \ref{spectral}.
Utilizing the previous results, spectral determinants
are identified   and the one missing parameter in section \ref{FRsl2}
will be determined.
The extension to arbitrary $n$ is the topics of sections \ref{FRsln}
and \ref{dvfsln}.
We conclude the paper with a brief summary and discussion in section \ref{sumdis}.

While preparing the manuscript, I find the preprint \cite{NewDT} appearing
on e-print.
The content of the paper 
largely overlaps with the second part of 
the present manuscript. They actually 
treated a more general set of
ODE but without the argument of the functional relations.

%
%
%-------------------------------------------------------------
%
%

\section{ Asymptotic Expansion, FFS and Stokes multipliers} \label{defStokes}
\noindent The details of the
present section can be found in  \cite{Sbook, HS, Fbook}.

We  first discuss the asymptotic behavior of a slightly generalized
differential equation,
\begin{eqnarray}
& &\partial^{n+1} y + (-1)^n P(x) y =0  \label{difp} \\
& & P(x) = \sum_{j=0}^{\ell} a_j x^{\ell-j} \np
\end{eqnarray}
where $a_j$ are complex numbers and $a_1=1$.
Note that the factor $(-1)^n$ is not essential.
It can be adsorbed into re-definition of the angle of $x$.
For later convenience, we will include this factor throughout out this report.

Now that $x=\infty$ is an irregular singular point of the equation,
 analytic properties of the solutions are different for 
different angle regions in complex $x$ plane.
Let ${\cal S}_k$ be a region in the plane satisfying
$$
|{\rm arg} x +k \theta  | \le \frac{\pi}{\ell+n+1}  
$$
for $ x \in {\cal S}_k$, where $\theta= \frac{2 \pi}{\ell+n+1}$.
We first analyze the asymptotic behavior of a subdominant solution in 
${\cal S}_0$.
Following \cite{HS,Sbook}, we define $b_h (h=1, 2, \cdots)$ by the relation,
$$
(1+ \sum_{k=1}^{\ell} a_k x^{-k}) ^{1/(n+1)} = 1+\sum_{h=1}^{\infty} b_h x^{-h}.
$$
A key function $E(x, {\bf a})$ is defined by $b_h$,
\begin{eqnarray*}
E(x,{\bf a}) &:=&
\int (1+\sum_{h=1}^{h_{\ell}} b_h x^{-h}) x^{\ell/(n+1)} dx  \\
        &=& \frac{n+1}{\ell+n+1} x^{(\ell+n+1)/(n+1)}+
		\sum_{h=1}^{h_\ell} \frac{b_h}{\frac{\ell}{n+1}-h+1} x^{\ell/(n+1)+1-h}
\end{eqnarray*}
where $h_{\ell} = N$ for $\ell=N(n+1) -j , (j=1, \cdots, n)$. 
${\bf a}$ stands for $(a_1,a_2, \cdots, a_{\ell})$

In addition, we introduce an exponent $\nu_{\ell}$ by
\begin{equation}
\nu_{\ell}=
\begin{array}{rl}
 \frac{n\ell}{2},&
           \quad \mbox{for $\ell \ne 0$ mod $n+1$}  \\
  \frac{n\ell}{2}+(n+1) b_{h_{\ell}+1},&
      \quad \mbox{for $\ell = 0$ mod $n+1$ }.\\
\end{array} 
\end{equation}
\begin{thm}
In  ${\cal S}_0$,  there exists a subdominant 
solution to (\ref{difp}) $y(x, {\bf a})$ 
which has the asymptotic behavior,

\begin{equation}
y(x, {\bf a}) \sim C^{-1} x^{-\nu_{\ell}/(n+1)} e^{-E(x,{\bf a})}.
\label{asy}
\end{equation}
\end{thm}

A normalization factor $C$ is introduced for convenience
in the later discussion,
$$
C^{n+1}:= \exp(-\frac{\pi n}{2} i) \prod_{0\le i <j\le n} (w^j-w^i), \quad 
w:=\exp(-\frac{2\pi}{n+1} i).
$$
As argued in \cite{SUNStokes}, the range of the validity of 
the asymptotic form is wider if one forgets the subdominance.
Explicitly, it is valid for $|{\rm arg } x| < \frac{n+2}{\ell+n+1} \pi$.

The intriguing feature in the differential equation (\ref{difp}) is
a certain symmetry in rotating $x$ plane.
\begin{thm} 
if $y(x, {\bf a})$ is the prescribed solution, then
$$
y_k(x, {\bf a}) :=
y(x q^{-k}, G^{(k)}({\bf a})) q^{n k/2}
$$
is also a solution to (\ref{difp}).
\end{thm}
The  parameter $q$ signifies $\exp(i\theta) =\exp(i\frac{2\pi}{\ell+n+1})$.
The operation $G^{(k)}({\bf a})$ is defined by 
$G^{(k)}({\bf a})=G(G^{(k-1)}({\bf a}) ),
k \ge 2$ 
and 
$G({\bf a})=(a_1/q, a_2/q^2, \cdots  a_{\ell}/q^{\ell})$.

From now on, we restrict our discussion to a single
potential term case , 
$$
P(x)= x^{\ell} + a_{\ell}, \quad  a_{\ell}=\lambda^{n+1}.
$$
One immediately verifies that  $b_{h_\ell+1}=0$ and thus 
$\nu_\ell=n\ell/2$ for $\ell>n+1$.
Under the operation of $G$, 
 $G(a_{\ell}) =a_{\ell} q^{-\ell} =a_{\ell} q^{n+1}$.
In term of $\lambda$,  the action of $G$ is simply given by
 $G^{(k)}(\lambda) =\lambda q^k$.
Consequently, $y_k= q^{nk/2}y(x q^{-k}, \lambda q^k)$.

A set of fundamental solutions (FSS) in ${\cal S}_k$ is formed by by
$(y_k, y_{k+1}, \cdots, y_{k+n})$.
We introduce a $(n+1) \times (n+1)$ matrix $\Phi_k(x)$
\begin{equation}
\Phi_k(x):=
\begin{pmatrix}
y_k,            & y_{k+1},           & \cdots,& y_{k+n}\\
\partial y_k,   & \partial y_{k+1},  & \cdots,&\partial y_{k+n}\\
\vdots          &                    &        &\vdots \\
\partial^n y_k, & \partial^n y_{k+1},& \cdots,&\partial y_{k+n}
\end{pmatrix}.
\end{equation}
We denote the Wronskian, the determinant of $\Phi_k(x)$, by $W_k$.
Note that the above asymptotic expansion is valid for $y_{k+j}, (j=0,\cdots,n)$
in the common sector ${\cal S}_{k+1/2} \cup {\cal S}_{k-1/2}$.
As $W_k$ is constant in $x$, 
one easily checks the linear independence of these solutions by
using the asymptotic expansion  (\ref{asy}) at the sector.
Due to the present normalization of $y_k$, we have $W_k=1$.

A  Stokes matrix $S_k$ connects FFS of ${\cal S}_{k}$
and ${\cal S}_{k+1}$

\begin{equation}
\Phi_{k+1}(x) = \Phi_{k}(x) S_k.
\label{stokes0}
\end{equation}

The linear independence of solutions demands $S_k$ in the following form,

\begin{equation}
S_k =
\begin{pmatrix}
\tau^{(1)}_1(\lambda q^k),  & 1, & 0,&  0,&  \cdots, & 0\\
\tau^{(2)}_1(\lambda q^k),  & 0, & 1,&  0,&  \cdots, & 0\\
      \vdots   &    &    &  &          & \vdots \\
\tau^{(n)}_1(\lambda q^k),  & 0, & 0,&  0,&  \cdots, & 1 \\
\tau^{(n+1)}_1(\lambda q^k),& 0, & 0,&  0,& \cdots,  & 0
\end{pmatrix}.
\label{sk0form}
\end{equation}
We call elements $\tau$ Stokes multipliers.

%The (1,1) component of (\ref{stokes}) is of later interest.
%\begin{equation}
%y_k = \tau^{(1)}_1(\lambda q^k) y_{k+1} + \tau^{(2)}_1(\lambda q^k) y_{k+2}+ 
%%
%\cdots + \tau^{(n+1)}_1(\lambda q^k) y_{k+n+1}
%\label{tq}
%\end{equation}

By the Cramer's formula, one represents $\tau^{(j)}_1(\lambda q^k)$ as

\begin{equation}
\tau^{(j)}_1(\lambda q^k)={\rm  det }
\begin{pmatrix}
y_{k+1},& y_{k+2}, &\cdots, & y_k, &  \cdots, & y_{k+n+1}\\
\vdots  &          &        &      &          & \vdots \\
\partial^{n} y_{k+1},&\partial^{n} y_{k+2}, &\cdots,& \partial^{n}y_k,& 
 \cdots, & \partial^{n}y_{k+n+1}
\end{pmatrix}
\label{wrep}
\end{equation}
that is, $(y_k, \partial y_k, \cdots, \partial^{n}y_k)$ is inserted
in the $j-$th column in the denominator.
Evidently  $\tau^{(n+1)}_1(\lambda q^k) = (-1)^n W_{k+1}/W_k=(-1)^n$.
 
The above representation (\ref{wrep}) of Stokes multipliers will be
referred to as the Wronskian representation.

Determinants of such structure will be hereafter abbreviated to, 
 by specifying only the first row,
$[y_{k+1}, y_{k+2}, \cdots,  y_k,   \cdots,  y_{k+n+1}]$.
Generally,
$$
[y_{i_1}, y_{i_2}, \cdots,    y_{i_n}]:=
\begin{pmatrix}
y_{i_1},& y_{i_2}, &  \cdots, & y_{i_n}\\
\vdots  &          &          &  \vdots \\
\partial^{n-1} y_{i_1},&\partial^{n-1} y_{i_2}, &
\cdots,&   \partial^{n-1}y_{i_n}
\end{pmatrix}.
$$

We have prepared materials needed for study on general $n$.
In the next few sections, however, we confine ourselves to the
$n=1$ case.
There are two reasons for the separated argument.
First, only for $n=1$ case, we have a clear bridge between
the connection problem and the spectral problem.
Second, the second order ODE 
may be the most relevant to physics.
%
%
%==============================================================
%
%

\section{ Fusion Stokes matrices for the 2nd order ODE }\label{FRsl2}

The ingenious idea in  \cite{DT2} lies in the
 introduction of  the generalized (or fusion)
Stokes matrices connecting the second neighboring sectors,
the third neighboring sectors, and so on.
We denote by $S^{(j)}_k $ the fusion Stokes matrices
connecting two FSS, $\Phi_{k}$ and $\Phi_{k+j}$

$$
\Phi_k = \Phi_{k+j} S^{(j)}_k.
$$

Obviously, the recursion relation holds,
\begin{equation}
S^{(j)}_{k} =  S^{(j-1)}_{k+1}  S^{(1)}_{k}.
\label{recur}
\end{equation}

\begin{thm}
$S^{(j)}_{k}$ has an expression
\begin{equation*}
S^{(j)}_{k} =
\begin{pmatrix}
         \tau^{(1)}_{j}(\lambda q^k),& \tau^{(1)}_{j-1}(\lambda q^{k+1}) \\
          - \tau^{(1)}_{j-1}(\lambda q^k),& -\tau^{(1)}_{j-2}(\lambda q^{k+1})
\end{pmatrix}
\end{equation*}
where we adopt $\tau^{(1)}_0(\lambda)=1,  \tau^{(1)}_{-1}(\lambda)=0$.
Thanks to the condition $y_{\ell+2+k}=- y_k$,
$\tau^{(1)}_{\ell}(\lambda)=-\tau^{(1)}_{\ell+2}(\lambda)=1$ and
 $\tau^{(1)}_{\ell+1}(\lambda)=0$.
Naturally, $\tau^{(1)}_{j}, (j \ge 2)$ are referred to as
the generalized Stokes multipliers.
Due to (\ref{recur}) they satisfy relations,
\begin{equation}
\tau^{(1)}_{j}(q \lambda)\tau^{(1)}_1(\lambda)\!=
\!\tau^{(1)}_{j+1}(\lambda)\!+\!\tau^{(1)}_{j-1}(q^{2}\lambda) .
\label{recursiontau}
\end{equation}
\end{thm}

\vskip 0.9cm 
\noindent {\bf example }\par
For $\ell=1$, 
$$
-\frac{d^2}{d x} y + x y = \lambda^2 y,
$$
it is well known that the eigenfunction is given by Airy function
$y={\rm  Ai}(x)$.
The above connection rule then fixes the Stokes multiplier for Airy function
$\tau^{(1)}_{1}=1$. \pn

\vskip 0.9cm 

Let $ \tau^{(1)}_j(\lambda)=  T^{(1)}_j (\lambda q^{(j+1)/2}) $.
One can then prove 
\begin{equation}
T^{(1)}_j(\lambda q^{1/2}) T^{(1)}_j(\lambda q^{-1/2})
= 1+
T^{(1)}_{j+1}(\lambda) T^{(1)}_{j-1}(\lambda).
\label{Tsyssl2}
\end{equation}
using  (\ref{recursiontau}) and the mathematical induction.
{\it These functional relations  exactly coincide with those among
fusion transfer matrices of} $U_q(A^{(1)}_1)$.
In the latter context,  the suffix
$j$ specifies the spin $j/2$ assigned to the auxiliary space.
They are the closed set of equations among finitely many unknown
functions $T^{(1)}_j, (j=0,1,\cdots, \ell)$. 
Thus they may be of significance in the estimation of the quantity of
our original interest, $\tau^{(1)}_1(\lambda)$.
Actually, with additional assumptions on the analyticity and
asymptotic behavior of  $\tau^{(1)}_j(\lambda)$, 
one can  fix  $\tau^{(1)}_1(\lambda)$ 
via coupled nonlinear integral equations resulting from ( \ref{Tsyssl2}).
To see this, we conveniently put 
$Y_{j}(\lambda)=T_{j+1}(\lambda) T_{j-1}(\lambda)$
and $\lambda={\rm e}^{\ell /(\ell+2) \theta }$.
Now the functional relations read
\begin{equation}
Y_j(\theta+i\frac{\pi}{h})  Y_j(\theta-i\frac{\pi}{h}) =
(1+ Y_{j-1}(\theta) ) (1+Y_{j+1}(\theta) ),
\quad j=1,2,\cdots, \ell-1
\label{Ysys}
\end{equation}
where $h=\ell$. Note that $Y_0=Y_{\ell}=0$.

\noindent {\bf Assumption} \label{assumeanzc}\pn
$(\log Y_j(\theta))'$  $((\log(1+  Y_j(\theta))') $
are analytic, nonzero and have constant asymptotic behavior (ANZC)
in the strips $ {\rm Im } \theta \in [-\frac{\pi}{h}, \frac{\pi}{h}]$,
${\rm Im } \theta \in [-0^+, 0^+]$  respectively.\pn
The validity of this assumption will be discussed in section \ref{spectral}.

Once this is granted, one immediately derives from (\ref{Ysys}) \cite{DT2,KP,JKSfusion},

\begin{equation}
\epsilon_j(\theta)= m_j r \exp \theta -
   \frac{1}{2\pi} \sum_{k=1}^{h-1} \phi_{j,k}*L_k(\theta)
 \label{sl2tba}
\end{equation}

where $m_j= \sin(\pi j/h)/ \sin(\pi/h)$, $Y_j(\theta)=\exp(\epsilon_j(\theta))$ and 
$L_j(\theta)=\log(1+1/Y_j(\theta))$.
The asterisk denotes the convolution, 
$A*B(\theta) = \int A(\theta-\theta') B(\theta') d\theta'$.

This type of coupled integral equations is known as
thermodynamic Bethe ansatz equation(TBA).
One finds them in various branches of IM, e.g., 
the thermodynamics of 1D spin chains or 
the perturbation theory of CFT.
The kernel, $\phi_{j,k}$, is related to the two particle S-matrix 
$S_{j,k}$
of quantum field theory based on $A_{h-1}$ 
by $\phi_{a,b}(\theta)=-i \partial_{\theta} \log S_{j,k}(\theta)$ and
\begin{eqnarray*}
S_{j,k}(\theta)&=&\prod_{j=0}^{{\rm  min}(j,k)-2}  \{|j-k|+2j+1 \}, \\
& & \{p\}:=(p-1)(p+1), 
  \quad (p)=\frac{\sinh(\theta/2+i \pi/2h)} {\sinh(\theta/2-i \pi/2h)}.
\end{eqnarray*}
\begin{thm}
The set of equations (\ref{sl2tba}) fixes $Y_j(\theta)$
for a given  $ r$.
\end{thm}
To determine the factor $r$, it needs an independent ingredient from the 
spectral theory.
We will come back to this point in section \ref{spectral}.

We have a remark.
For the later use in the spectral problem, we have introduced
generalized Stokes matrices and derived 
 functional relations (\ref{Tsyssl2}) from the obvious recursion relation
 (\ref{recur}).
They can be also easily  extracted from the 
following Wronskian representation of $\tau^{(1)}_j(E)$,

\begin{equation}
\tau^{(1)}_j(E) = {\rm det} 
\begin{pmatrix}
y_0,&  y_{j+1} \\
y'_0,&  y'_{j+1} \\
\end{pmatrix}.
\label{wronsk}
\end{equation}

For $n>1$, the situation is different. 
The generalized Stokes matrices can be defined similarly.
Their elements, however, do not contain nice
generalization of $\tau^{(a)}_1$'s.
The formal definition of the Wronskian type like (\ref{wronsk})
still works efficiently. See the discussion in section \ref{FRsln}.
%
%
%----------------------------------------------------
%
%
\section{Dressed Vacuum Forms of Stokes multipliers}\label{dvf}

As shown in the previous section, the Stokes multipliers
share same functional relations with the transfer matrices of 
IM.
Below we will discuss if this correspondence carries forward.

The eigenvalues of the transfer matrices in solvable models 
exhibit a universal structure often referred to as the dressed vacuum form (DVF).
We shall explain DVF for the simplest  the $A^{(1)}_1$ case
with the dimension of the auxiliary space being 2.

Obviously, the highest weight state (= vacuum) is 
the trivial eigenstate of the transfer matrix.
Its eigenvalue consists of two terms, reflecting
the dimensionality of the auxiliary space.
Each of them is given by the simple product of the local weights 
which is termed as the vacuum expectation value,
$$
T_{\rm vacuum } (\lambda) =f_1(\lambda)+ f_2(\lambda).
$$
This expression must be modified for general eigenvalues.
The quantum inverse scattering method yields the exact expression.
The result tells that $T_{\rm vacuum } (\lambda)$ must be modified by
"dressing " the  vacuum expectation values with ratios of Baxter's 
$Q$ operator (or its eigenvalue) which commutes with $T$, $[T, Q]=0$, 
\begin{equation}
T^{(1)}_1(\lambda) =f_1(\lambda) \frac{Q(\lambda q)}{Q(\lambda)}+ 
                  f_2(\lambda)  \frac{Q(\lambda q^{-1})}{Q(\lambda)}.
				  \label{DVF6v}
\end{equation}

The fact that the eigenvalue must be pole free results the famous
Bethe ansatz equation (BAE),
$$
 \frac{f_1(\lambda_j)}{f_2(\lambda_j)}=-
 \frac{Q(\lambda_j  q^{-1})}{ Q(\lambda_j  q ) }
$$
where $Q(\lambda_j)=0$.
This kind of representation is called DVF.

Clearly, eq( \ref{DVF6v}) has an interpretation 
as the second order difference equation (Baxter's T-Q relation),
$$
T^{(1)}_1(\lambda) Q(\lambda) = 
f_1(\lambda) Q(\lambda q) +  f_2(\lambda) Q(\lambda q^{-1}).
$$
Thus two independent solutions exist which we call $Q_{\pm}$.

In \cite{Reshet, KSanaly}, it is shown that DVF is  universal for models
based on general $U_q(\widehat{g})$ 
under certain assumptions. 
The key ingredient in the argument is the analyticity under BAE.
Thus we may conclude
that DVF embodies the BAE or Yang-Baxter integrable structure.

Now we turn to the  $n=1, k=0$ of (\ref{stokes0}).
The Stokes multiplier $\tau^{(1)}_1(\lambda)$ is given by $y's$ in two manners,

\begin{equation}
\tau^{(1)}_1(\lambda) = \frac{y_0+y_2}{y_1} |_{x=0}  \quad
                     =\frac{y'_0+y'_2}{y'_1} |_{x=0}.	 \label{tauDVF} 				  
\end{equation} 

Originally, the rhs can be evaluated at any $x$ yielding the same
$\tau^{(1)}_1(\lambda)$.
We adopt a convention to enumerate them at the origin for
the later convenience. See section \ref{spectral}.

By comparison of (\ref{DVF6v}) and  (\ref{tauDVF})
and the identification, $\tau^{(1)}_1(\lambda)=T^{(1)}_1(\lambda q)$,
made after (\ref{recursiontau})
, we deduce
$y_j \propto Q_-(\lambda q^{j})$ and 
 $y'_j \propto Q_+(\lambda q^{j})$.
Precisely, the argument in the next section concludes
$
y_j = q^{j/2} Q_-(\lambda q^{j}),  
$ and
$
y'_j = q^{-j/2} Q_+(\lambda q^{j}).
$

The linear independence of FSS implies that $\tau^{(1)}_1(\lambda)$
is pole-free.
On the other hand, $y(0, \lambda)$ can generally be zero
for some $\lambda=\lambda_j$.
Thus we have  BAE for Stokes multipliers.

It is interesting that $dy/dx$, which is by no means a solution
to the original ODE, now appears as the second "solution" to the
difference equation.
This issue will be further pursued in a later section.

The coincidence is not only for the spin 1/2 case,
but also for cases of arbitrary  spins.
This can be easily seen as they share the same initial condition
and the functional relations.
One can also verify this directly using the Wronskian representation.
For this we rewrite the condition $W_k=1$ in the form,

$$
y_k y'_{k+1} -y'_k y_{k+1}=1 \Longrightarrow
 \frac{y'_{k+1}}{y_{k+1}} - \frac{y'_k}{y_k} = \frac{1}{y_k y_{k+1}}.
$$

With use of this,  one obtains 

\begin{eqnarray*}
\tau^{(1)}_j(\lambda) & =& 
y_0 y_{j+1} (\frac{y'_{j+1}}{y_{j+1}}-\frac{y'_{0}}{y_{0}})
=y_0 y_{j+1} \sum_{k=1}^{j}   
 (\frac{y'_{j+1-k}}{y_{j+1-k}}-\frac{y'_{j-k}}{y_{j-k}}) \\
&=& y_0 y_{j+1} \sum_{k=1}^{j} \frac{1}{y_{j-k} y_{j-k+1}},
\end{eqnarray*}

which coincides with the known expression for the
transfer matrix.
Actually the discussion like above has been firstly found as
the operator identity under the name of the quantum Wronskian form. 
We follow the discussion in \cite{BLZ2, BLZ3} 
for reproducing the DVF for the spin $j/2$ case.

Before closing the section, we present simplest examples
($\ell=1,2$ ) where explicit solutions
are available by elementary functions \cite{ DT1, DT2, BLZnew, V5,V8}.
We shall use $E$ instead of $\lambda$ ($E=\lambda^2$)
and adopt same symbols, y, $\tau$ etc as the 
function of $E$. \pn
\vskip 0.9cm
\underline{The  case  $\ell=1$} \pn
This is a well known example in  quantum mechanics.
The wave function $y$ is given by the Airy function.

On the other hand, for $\ell=1$, we have $\tau^{(1)}_1=1$ (theorem 3).
Thus $T-Q$ relation simplifies,

\begin{eqnarray}
Q_-(E) &=& q^{-1/2} Q_-(E q^{-2}) + q^{1/2} Q_-(E q^2)  \nonumber \\
Q_+(E) &=& q^{1/2} Q_+(E q^{-2} ) + q^{-1/2} Q_+(E q^2).
\label{simpletq}
\end{eqnarray}

These relations coincide with the 3-solution dependence
relation for the Airy function\cite{V5},

\begin{eqnarray}
q^{-1} {\rm Ai}(q^{-1} E) +  {\rm Ai}( E) +  q {\rm Ai}(q  E)&=&0  \nonumber\\
q^{-2} {\rm Bi}(q^{-1} E) + {\rm Bi}( E) +  q^2 {\rm Bi}(q E)&=&0 
\label{airyrel}
\end{eqnarray}
where ${\rm Bi}(x):= \frac{d {\rm Ai}(x) }{dx}$.

To check this,  we use $q^3=1$ in the arguments in the first
of (\ref{simpletq}),
$$
Q_-(E) = q^{-1/2} Q_-(E q) + q^{1/2} Q_-(E q^{-1}),
$$
and substitute $q=-q^{-1/2}$ in the coefficients of
the first relation in (\ref{airyrel}),
$$
-q^{1/2} {\rm Ai}(q^{-1} E) +  {\rm Ai}( E) -  q^{-1/2} {\rm Ai}(q  E)=0,
\Longrightarrow 
{\rm Ai}( E) = q^{-1/2} {\rm Ai}(q  E)+ q^{1/2} {\rm Ai}(q^{-1} E).
$$
The second relations can be checked similarly.
\vskip 0.9cm
\underline{The  case  $\ell=2$} \pn
The case with the harmonic oscillator is 
slightly complicated as the asymptotic formula
must be modified.
We utilize  known facts on the Weber's function $D_{\eta}(z)$,
$$
\frac{d^2 D_{\eta}(z)}{d z^2} +(\eta+\frac{1}{2}-\frac{z^2}{4})D_{\eta}(z) =0,
$$
which has an asymptotic behavior for $\eta \ne 0,  {\rm integer }$,
$$
D_{\eta}(z) \sim z^{\eta} \exp(-z^2/4).
$$
It has the 2nd order irregular singularity 
at $\infty$ and regular elsewhere.

The FSS consists of  $ \{ D_{\eta}(z), D_{-\eta-1}(iz)  \}$ or
$ \{ D_{\eta}(-z), D_{-\eta-1}(-iz)  \}$.
The connection rule reads,

\begin{equation}
\frac{\sqrt{2\pi}}{\Gamma(\eta+1)} D_{\eta}(z) =
i^{\eta} D_{-\eta-1}(iz) +  i^{-\eta} D_{-\eta-1}(-iz).
\label{weberconn}
\end{equation}

There exist recurrence relations,
\begin{equation}
D'_{\eta}(z) = z/2 D_{\eta}(z) -D_{\eta+1}(z) =
 -z/2 D_{\eta}(z) +\eta D_{\eta-1}(z).
\label{recurd}
\end{equation}

Obviously, $y(x, E)$ is given in terms of $D_{\eta}(z)$.
In order to cancel the phase factor arising from
the asymptotic behavior, we define precisely
$$
y_k:= q^{k/2 +k/2 E_k } 
\frac{ D_{\eta_k}(\sqrt{2} x q^{-k}) }{ 2^{\eta_k /2} \sqrt{2 i} }
$$
where $E_k= E q^{2 k}$, $2 \eta_k+1=E_k$ and $\eta=\eta_0$.
They constitute our FSS.

By definition, $T^{(1)}_1(E) y_0=y_1+y_{-1}$. 
Remembering  $q=i$ so that $\eta_{1}=\eta_{-1}=-\eta-1$
, we rewrite this into the form,
$$
\frac{T^{(1)}_1(E)}{2^{\eta+1/2}} D_{\eta}(z) =
i^{\eta} D_{-1-\eta}(iz)+ i^{-\eta} D_{-1-\eta}(-iz),
$$
with $z=\sqrt{2} x$.
By comparing this with (\ref{weberconn}) we conclude
$$
T^{(1)}_1(E)= 2^{\eta+1} \frac{\sqrt{\pi}}{\Gamma(\eta+1)}
 = 2^{E/2+1/2}  \frac{\sqrt{\pi}}{\Gamma(E/2+1/2)}
$$
which coincides with the result from CFT\cite{DT2, BLZnew}.
The expectation values of $Q_{\pm}(E)$ are proportional to Weber's function
and its derivative at the origin.
Thanks to the recursion relations (\ref{recurd}), we can replace the latter 
by again Weber's function with the unit shift in $\eta$,

$$
Q_-(E) \propto D_{\eta}(0), \qquad  Q_+(E) \propto D_{\eta+1}(0).
$$ 

We shall utilize the following integral representation for 
$D_{\eta}(z)$,
$$
D_{\eta}(z) =-\frac{ \Gamma(\eta+1)}{2\pi i} {\rm e}^{-z^2/4}
 \int_{\cal C} {\rm e}^{-t^2/2 -zt} (-t)^{-(\eta+1)} dt,
$$
where ${\cal C}$ surrounds the positive real axis counterclockwise.
The evaluation at $z=0$ is then straightforward,
$$
D_{\eta}(0) = \frac{2^{\eta/2} \sqrt{\pi}}{\Gamma((1-\eta)/2)}
=\frac{2^{(E-1)/4} \sqrt{\pi}}{\Gamma((3-E)/2)}.
$$
Hence,

\begin{equation}
Q_-(E) \propto \frac{1}{\Gamma((3-E)/4)},
\qquad  Q_+(E) \propto \frac{1}{\Gamma((1-E)/4)}.
\label{qforl2}
\end{equation}

For general values of $\ell$, the representations of
$T^{(1)}_1$ or $Q_{\pm}$ by elementary functions are not
known.
Still, we can evaluate them, e.g., from solutions to
TBA (\ref{sl2tba}).
To fix one missing parameter $r$ there, we next consider
the spectral problem.

%
%============================================================
%
%

\section{Spectral Determinants  and Stokes multipliers}\label{spectral}

The final section for the $n=1$ case is devoted to the 
spectral problem for $\ell= 2M$ and $M$ being an integer.
We will still use $E$ instead of $\lambda$. 

We first put some remarks on elementary facts.
Let $H(x)$ be a our Hamiltonian operator,
$H(x) = -\frac{d^2}{ dx^2} + x^{2M}$.

\begin{df}
We call $\psi(x)$ the eigen-function and $E$, the eigenvalue of $H(x)$
if  $H(x) \psi_E(x) = E \psi_E(x)$ and $ \psi_E(x) $ is a vector
in the Hilbert space satisfying, e.g., $ ||\psi(x)||<\infty $.
\end{df}

\begin{df}
Let $P$ be a spatial inversion operator such that 
$P f(x) = f(-x)$ for any operators or vectors.
\end{df}

Obviously, $[H(x), P]=0$.
Thus if $H(x) \psi_E(x) = E \psi_E(x)$ then $P\psi_E(x)=p\psi_E(x)$.
Since $P^2$  is an identity operator, $P\psi_E(x)= \psi_E(-x)=\pm \psi_E(x)$.
Consequentially, we have
\begin{lem}
If $H(x) \psi_E(x) = E \psi_E(x)$  then
$\psi_E(x=0)=0$ or $\frac{d \psi_E(x)}{dx}|_{x=0}=0$.
\end{lem}
The above lemma does not require 
the boundary condition $ \lim_{x \rightarrow \pm \infty} |\psi_E(x)| =0 $
imposed by our potential. 

Two conditions can not be satisfied simultaneously.
Or otherwise, $\psi_E^{(n)}(x=0)=0$ for arbitrary $n$, resulting
a trivial $\psi$.
Thus a lemma follows.
\begin{lem}
 Eigenvalues are classified by the parities of the associated 
eigenfunctions.
We denote $E^{+}_j $ if $\frac{d \psi_{E^+_j}(x)}{dx}|_{x=0}=0$
and  $E^{-}_j $ if $\psi_{E^-_j}(x=0)=0$
\end{lem}

On the positive real axis,  $\psi_E(x)= y_0 + a(E) y_1$
up to normalizations.
We have $[\psi_E(x), y_0]=a(E)$ from the obvious asymptotic behavior, 
$\lim_{x \rightarrow \infty} \psi_E(x)/ y_1= a(E)$.
As the eigenfunction must be bounded, $a(E)=0$ if 
$E \in  \{ E^{+}_{j}   \} \cup \{ E^{-}_{j}  \}   $.
Conversely, if $a(E')=0$ for some $E'$ then $\psi_{E'}(x)$ is 
proportional to  $y_0$.
Thus  $\psi_{E'}(x)$  is bounded as $x \rightarrow +\infty$
{\it and } it is recessive as $x \rightarrow -\infty$ due to
the parity argument.
In addition, it is a solution to the eigenvalue 
equation.  Then, by definition, $E'$ belongs to the set of eigenvalues.
We conclude, 
\begin{lem}\label{de}
If  $\psi_E(x)= y_0 + a(E) y_1$ on the positive real axis,
then $a(E) \propto D(E):=D_+(E) D_-(E)$
where  $ D_{\pm}(E):= \prod_j (1- E/E^{\pm}_{j})$.
\end{lem}

Finally we quote results from the WKB analysis,

\begin{lem}
For the potential $x^{2M}$, the energy levels $E_k$ and
the spectral determinant $D(E)$  behave asymptotically as
\begin{eqnarray}
b_0 (E_k)^{\mu} &\sim&  2\pi (k+1/2), \quad k \rightarrow \infty  \\
\ln D(E) &\sim&  \frac{b_0}{2 \sin(\mu \pi)} E^{\mu} \\
\mu&=& \frac{M+1}{2M}, \quad
b_0=\frac{\pi^{1/2} \Gamma(\frac{1}{2M}) }
{M  \Gamma(\frac{1}{2M}+\frac{3}{2})} .
\end{eqnarray}

\end{lem}

We shall apply the above general observation to results obtained
in the preceding sections.
The connection rule  enables a representation of $D(E)$ in terms of $y_0$.
To check this, we consider the Stokes matrix $S^{(M+1)}_0$.
It connects FSS on the positive and the negative real axes.
We start from the  negative real axis.
If $E$ takes an eigenvalue, then 
$\psi_{E}(x)= y_{M+1}$  apart from a normalization.
The connection rule demands it behave on the 
positive axis,
$$
 \psi(x, E_{\alpha}) \sim -\tau^{(1)}_{M-1} (q^2 E_{\alpha}) y_0 +
 \tau^{(1)}_M(E_{\alpha}) y_1.
$$
Lemma \ref{de} tells  $\tau^{(1)}_M \propto  D(E)$.

On the other hand, we consider the $j=M$ case  of (\ref{wronsk}).
Note that  $q^{M+1}=-1$ , and $y_{M+1} = i y(-x, E)$.
Then  $\tau^{(1)}_M=i(y(x,E) y'(-x, E)- y(-x,E) y'(x, E))$.
Since the lhs is independent of $x$ and the rhs is not singular at $x=0$,
we conveniently put $x=0$ in the rhs and find 
$\tau^{(1)}_M \propto y(0,E) y'(0,E) |_{x=0}$. 
Thus, as a function of $E$, $y_0 y'_0 |_{x=0}$ has only zeros at
eigenvalues. 
Then the above lemma leads to their identification with $D^{\pm}(E)$.
The choice of the evaluation at $x=0$ here and in the previous section
is now clear.

Summarizing, we have a theorem.

\begin{thm}\label{thde}
The fusion hierarchy contains $D(E)$ as its $M-$ th member,
$$
\tau^{(1)}_M \propto  D(E), \quad {\rm equivalently } \quad
T^{(1)}_M(E) \propto  D(-E). 
$$ 
A base of FSS and its derivative at the origin are proportional to
spectral determinants depending on parities,
$$
y(0, E) \propto D_-(E), \quad y'(0,E) \propto D_+(E).
$$
\end{thm} 

The previous explicit result (\ref{qforl2}) for $M=1 (\ell=2) $ 
 is quite consistent  with this.
 $Q_{\pm}(E)$'s are nothing but $D_{\pm}(E)$ here.
They are vanishing at known spectra of the harmonic oscillator,
$E=2 n+1$, where $n=$(even/odd) corresponds to the parity $=$ (even/odd).

These identifications lead to the expression for $T^{(1)}_1$ via 
spectral determinants,
$$
T^{(1)}_1(E) = q^{1/2} \frac{D_+(E q^2)}{D_+(E)}
+ q^{-1/2} \frac{D_+(E q^{-2})}{D_+(E)}
= q^{-1/2} \frac{D_-(E q^2)}{D_-(E)}
+  q^{1/2} \frac{D_-(E q^{-2})}{D_-(E)}.
$$
More significantly, we have BAEs,
$$
 \frac{ D_{\epsilon}(E_j q^2)}{ D_{\epsilon}(E_j q^{-2})} =
 -q^{\epsilon}, \qquad \epsilon=\pm.
$$
These equations, combined  with the WKB result, 
are efficient enough to determine the spectral determinants, 
being transformed into  coupled nonlinear integral equations.
We, however, take a different route here and utilize them as a tool
to investigate TBA (\ref{sl2tba}).

Let us revisit to the assumption \ref{assumeanzc} raised in section \ref{FRsl2}.
Suppose all energy levels are enumerated exactly so that 
 $D_{\pm}(E)$ are constructed.
Then $T^{(1)}_1(E)$ is estimated. 
By the use of the analogue of the relation(\ref{recursiontau}), 
we can successively generate
$T^{(1)}_j(E)$ and check the validity of the assumption.
Strictly speaking, as we have infinitely many levels, this procedure 
can not be accomplished.
One however knows that the WKB approximation is fairy accurate for 
higher energy levels. 
 Thus we input first 100 exact energy levels
and approximate rests by the WKB results, to evaluate 
$D_{\pm}(E)$.

\begin{figure}
\centering
  \includegraphics[width=6cm]{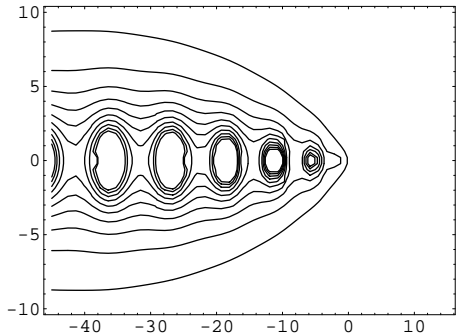}
\caption{the contourplot for $M=3$, 
$| {\rm e}^{E/7} T_2^{(1)}(E) |$}
\label{figzeros}
\end{figure}

Our numerical results indicate the  remarkable patterns,
\begin{cj}
Zeros of $T^{(1)}_j(E)$  are of the first order and always distribute on the
negative real $E$ axis.
\end{cj}
This supports the assumption, although by no means a proof.
As an example,
the contourplot for $M=3$, 
$| {\rm e}^{E/7} T_2^{(1)}(E) |$ is depicted in Fig.\ref{figzeros}.
These patterns imply that the state
corresponds to "the vacuum" (the ground state) in IM \cite{DT1,BLZnew}.

There is also an independent support to this conjecture \cite{DT2,PT1,PT2}.
As shown in  \cite{DT2},
zeros of $T^{(1)}_j(E)$ coincide with negative of eigenvalues
associated to ${\cal P}{\cal T}$-symmetric Hamiltonian 
$p^2 + x^{2j} (ix)^{\epsilon}$ with $\epsilon=2M-2j$.
The  numerical and analytical studies on the ${\cal P}{\cal T}$-symmetric Hamiltonian 
in \cite{PT1,PT2} conclude
positive and real eigenvalues for $M \ge 1$, which is consistent with 
the conjecture.
The studies reveal, at the same time,  the breakdown of the conjecture,
when $M$ being continued to a real number less than 1 \cite{DT2,PT1,PT2}.

We assume that the validity of integral equations (\ref{sl2tba}).
Then all we have to do is fix $r$
in evaluating Stokes multipliers and spectral determinants.
The theorem \ref{thde} tells $T^{(1)}_M(-E_k)=0$, which 
results $Y_M(-E_k q^2) = -1$ or 
$\log Y_M(\theta_k+i\frac{\pi}{2})=(2 k+1) \pi i$.
Remember $E=\exp(\theta/\mu)$.
For large values of $\theta$, numerical data concludes
that the contribution form the integral is negligible
so that we have an approximation,
$
\log Y_M(\theta) \sim m_M \exp(\theta).
$
Thus $T^{(1)}_M(-E_j)=0$ means 
$
m_M \exp(\theta_k)= m_M E_k^{\mu}=(2k+1) \pi
$
 for large enough $j$.
Comparing this with the WKB result, we conclude 
$m_M r=b_0$, which derives the desired quantity.

Summarizing the results for 2nd order ODE, we have the
following correspondence,

\vskip 0.9cm
\fbox{
\begin{tabular}{rrl}
energy &    $\Longleftrightarrow$ &     spectral parameter \\
Stokes multipliers, $D(E)$&  $\Longleftrightarrow$ &  fusion transfer matrices \\
$y|_{x=0}, y'|_{x=0}$ &  $\Longleftrightarrow$ &    
   (vacuum) expectation values of $Q_{\pm}$  \\
\end{tabular}
}
\vskip 0.9cm

In the next two sections, the higher order ODE
will be briefly discussed.
Our results indicate a natural extension of
the ODE/IM correspondence examined for $n=1$ above.

%
%
%-------------------------------------------------------------
%
%

\section{Functional relations in Stokes multipliers}\label{FRsln}

As in the case of $n=1$, we can introduce 
generalized Stokes
matrices connecting disjoint sectors for arbitrary $n$.
The obvious recursion relation leads to functional relations, %
however, among complex objects corresponding to 
Young tableaux of the hook shape.
Then the restriction of relations among Young tableaux of the rectangular shape
results the desired relation \cite{SUNStokes}.
This procedure requires some technique in integrable models
e.g., quantum analogue of the Jacobi-Trudi formula.

We derive the same relation in a simpler way using the Wronskian
representation of Stokes multipliers.
Let auxiliary functions $\tau^{(a)}_m(\lambda)$ be

%\begin{equation}
%\tau^{(a)}_m (\lambda)= {\rm  det }
%\begin{pmatrix}
%y_1          &     \cdots      & y_{a-1}          & 
%y_0          &      y_{a+m}    &    \cdots& y_{n+m-1}\\
%  
%\partial y_1 &         \cdots  & \partial y_{a-1} & 
%\partial y_0 &  \partial y_{a+m}&   \cdots&  \partial y_{n+m-1}\\
%
%\vdots       &                 &                  & 
%             &                 &          &\vdots \\
%\partial^{n}y_{1}&      \cdots & \partial^{n} y_{a-1}&
% \partial^{n} y_{0}& \partial^{n} y_{a+m}& \cdots & \partial^{n}y_{m+n-1}\\
%\end{pmatrix}
%\label{tauamform}
%\end{equation}
%

\begin{equation}
\tau^{(a)}_m (\lambda)= [
y_1, y_2, \cdots  y_{a-1},y_0 , y_{a+m},y_{a+m+1}\cdots y_{n+m} ].
\label{tauamform}
\end{equation}
Note that we adopt the abbreviation defined in section 2.

Due to $y_{n+1+\ell}=(-)^n y_{0}$, 
 $\tau^{(a)}_m(\lambda)=0 $ for $m \ge \ell+1, \ell+2,\cdots$. 
This is an analogue to the quantum group reduction.
Remark that the set contains the original Stokes multipliers as $m=1$ cases.

Then the claim is the following functional relations , 
\begin{thm}\label{Tsystem}
\begin{equation}
\tau^{(a)}_m(\lambda) \tau^{(a)}_m(\lambda q)
= \tau^{(a+1)}_m(\lambda) \tau^{(a-1)}_m(\lambda q)+
 \tau^{(a)}_{m+1}(\lambda) \tau^{(a)}_{m-1}(\lambda q)
\label{Tsys}
\end{equation}
where $\tau^{(0)}_1=1,  \tau^{(a)}_0 = (-1)^{a-1}$.
\end{thm}

We utilize a lemma in \cite{Hirota} 
for the proof of the above theorem.

\begin{lem}
\begin{eqnarray}
& &[f_1,f_2, \cdots f_N, a_0,a_1][f_1,f_2, \cdots f_N,a_2,a_3]- \\
& & [f_1,f_2, \cdots f_N,a_0,a_2][f_1,f_2, \cdots f_N,a_1,a_3]+ \\
& & [f_1,f_2, \cdots f_N,a_0,a_3][f_1,f_2, \cdots f_N,a_1,a_2]=0 .
\label{hirota}
 \end{eqnarray}
\end{lem}

When $N=1$,
this relation follows from the Laplace expansion of the trivial relation,
$$
0={\rm det }
\begin{pmatrix}
f    &   a_0   &  a_1   &  0  &  a_2 & a_3   \\
f'   &   a'_0  &  a'_1  &  0  & a'_2& a'_3  \\
f"   &   a"_0  &  a"_1  &  0  & a"_2& a"_3  \\
0    &      0  &  a_1   &  f  & a_2 & a_3   \\
0    &      0  &  a'_1  &  f' & a'_2& a'_3  \\
0    &      0  &  a"_1  &  f" & a"_2& a"_3  
\end{pmatrix}.
$$
With  the similar argument,
the validity of the above lemma is verified for arbitrary $N$.

\begin{proof}
\noindent of theorem \ref{Tsystem} : \pn
We shall adopt  identifications,

\begin{eqnarray*}
(f_1, f_2, \cdots, f_N) &\leftrightarrow& 
(y_1, \cdots, y_{a-1}, y_{a+m+1},  \cdots,y_{n+m}) \\
(a_0, a_1, a_2, a_3)  &\leftrightarrow&  (y_0, y_a, y_{a+m}, y_{n+m+1}).
 \end{eqnarray*}
 
For $a=1$, the left hand side of the first relation should read
as $(y_2, \cdots, y_{n+1})$.
The six elements in eq.(\ref{hirota}) are interpreted as
\begin{equation}
\begin{matrix}
 -\tau^{(a+1)}_m(\lambda),& (-1)^{n-2} \tau_m^{(a-1)}  (\lambda q), \\
  \tau^{(a)}_m(\lambda),& (-1)^{n-1} \tau_m^{(a)} (\lambda q), \\
 (-1)^{n-a} \tau_{m+1}^{(a)}  (\lambda),&
                           (-1)^{a-1} \tau_{m-1}^{(a)}  (\lambda q)\\
\end{matrix}
\nonumber 
 \end{equation}
respectively.
This immediately leads to the theorem. 
\end{proof}

We have a note.
The substitutions $\tau^{(a)}_m(\lambda) \rightarrow 
(-1)^{a-1} T^{(a)}_m(\lambda q^{(a+m)/2})$
brings the equation to the form known as the $T-$ system for $A^{(1)}_n$ 
in solvable models,
\begin{equation}
T^{(a)}_m(\lambda q^{1/2})  T^{(a)}_m(\lambda q^{-1/2}) = 
T^{(a+1)}_m(\lambda )  T^{(a-1)}_m(\lambda ) +
T^{(a)}_{m+1} (\lambda )  T^{(a)}_{m-1} (\lambda ).
\label{kns}
\end{equation}
This observation supports the ODE/IM correspondence
 for $n$ arbitrary.
In view of the solvable models, $T^{(a)}_m(\lambda)$ should be
understood as the (eigenvalues of) transfer matrix associated to
the auxiliary space $W^{(a)}_m(\lambda)$. ( As the module in 
classical Lie algebra, $W^{(a)}_m(\lambda)$ is isomorphic to
$m \Lambda_a$, of which Young diagram takes a rectangular shape.)

The relation also finds a connection to a discrete soliton system.
We parameterize $\lambda= q^{p/2} $ and denote
$f(p,a,m) =T^{(a)}_m(\lambda)$.
Let $D_i, i=1,2,3$ be Hirota operators acting on $i$ th variable.
Then the eq (\ref{kns}) reads 
$$
(\exp D_1 - \exp D_2 - \exp D_3) f \cdot f =0.
$$
This equation is known as the Hirota-Miwa equation with $Z_1=-Z_2=-Z_3=1$
\cite{Hiro,Miwa}.
The present construction imposes the periodicity and boundary conditions,
\begin{eqnarray*}
f(\ell+n+1,a,m) &=&f(0,a,m) \\
f(p,-1,m)&=& f(p,n+2, m)=f(p,n+1,\ell+1) =0.
\end{eqnarray*}

%
%
%===============================================================
%
%
\section{DVF for arbitrary $n$   }\label{dvfsln}

The DVF in the Stokes multipliers are also found for arbitrary $n$.  

Let us check this for $\tau^{(1)}_1(\lambda)$ .
The following lemma is useful for this purpose.

\begin{lem}\label{recursiondvf}
For $m \ge 2$ we have a recursion relation among ratios of
determinants,
\begin{equation}
\frac{[y_0, y_2, \cdots y_m] }{ [ y_1, \cdots ,y_m  ]}=
\frac{[y_0, y_2, \cdots y_{m-1}] }{ [ y_1, \cdots ,y_{m-1} ]}+
\frac{[y_0, y_1, \cdots y_{m-1}][y_2, \cdots, y_m] }
  { [ y_1, \cdots ,y_m ][ y_1, \cdots ,y_{m-1} ]}.
  \label{recusion}
\end{equation}
We should interpret $[y_1] \rightarrow 
y_1$ and $[y_0, y_1, \cdots y_{m-1}] \rightarrow  y_0$ for $m=2$.
\end{lem}

\begin{proof} \pn
The lemma is equivalent to

\begin{equation}
[y_0, y_2, \cdots y_m] [ y_1, \cdots ,y_{m-1} ]=
[y_0, y_2, \cdots y_{m-1}] [y_1, \cdots, y_m]+
[y_0, y_1, \cdots y_{m-1}][y_2, \cdots, y_m] .
  \label{recusion2}
\end{equation}

Since (\ref {recusion2}) is linear in $y_0^{(j)}$,
 it suffices to
show the equality of the coefficients of them in the both sides.
First consider the coefficient of $y_0$,
We need to show the equality,
\begin{equation}
[y_1, \cdots, y_{m-1}][y'_2, \cdots y'_m] =
[y_2, \cdots, y_{m}][y'_1, \cdots y'_{m-1}]+
[y_1, \cdots, y_{m}][y'_2, \cdots y'_{m-1}].
\label{plucker}
\end{equation}

To verify this, we prepare a matrix ${\cal M}$
$$
{\cal M}:= 
\begin{pmatrix}
  y_1,&            y_2&    \cdots& y_m \\
 y'_1,&           y'_2&   \cdots& y'_m \\
 \vdots&               &   \cdots&   \vdots \\
y^{(m-1)}_1& y^{(m-1)}_2& \cdots&   y^{(m-1)}_m\\
\end{pmatrix}.
$$
Denote by 
$$
D \begin{bmatrix} i_1,&i_2,&\cdots \\   j_1,&j_2,&\cdots \\     \end{bmatrix}
$$
the minor, the determinant of a matrix obtained
 by deleting $i_1 , i_2 \cdots$ rows and 
$j_1, j_2 , \cdots$ columns from ${\cal M}$.
Then eq.(\ref{plucker}) is represented as,
$$
D \begin{bmatrix} m \\  m \\  \end{bmatrix} D \begin{bmatrix} 1 \\  1 \\  \end{bmatrix}
=
D \begin{bmatrix} 1 \\  m \\  \end{bmatrix} D \begin{bmatrix} m \\  1 \\  \end{bmatrix}
+ D  D \begin{bmatrix} 1,&m \\   1,&m \\   \end{bmatrix},
$$
where $D={\rm det}{\cal M}$.
Obviously this is the Jacobi identity.
Thus the equality of coefficients of $y_0$ in both sides is established.
The equalities are similarly proven up to those of $y_0^{(m-2)}$.
For $y_0^{(m-1)}$ case, the first term of the rhs 
in (\ref{recusion2}) does not contribute.
We can check , however, the equality of the reminding terms.
Thus the lemma is proved.
\end{proof}

Next we will show

\begin{thm}
$\tau^{(1)}_1(\lambda) $ can be represented in
the following DVF
\begin{eqnarray}
\tau^{(1)}_1(\lambda) &=& 
\frac{[y_2, y_3, \cdots, y_{n+1}]}{[y_1,y_2,\cdots, y_n]}+
\frac{[y_2, y_3, \cdots, y_{n}] [y_0, y_1, \cdots, y_{n-1}]}
      {[y_1,y_2,\cdots, y_{n-1}] [y_1,y_2,\cdots, y_{n}]} +  \nonumber  \\
& & \cdots+ \frac{[y_2,y_3] [y_0, y_1,y_2]}{[y_1,y_2] [y_1,y_2,y_3]} + 
  \frac{[y_0, y_1] y_2}{[y_1,y_2] y_1} +\frac{y_0}{ y_1}.
\label{DVFsln}
\end{eqnarray}
\end{thm}

\begin{proof} \pn
Firstly we substitute $\tau^{(n+1)}_1=(-1)^n $ to eq.(\ref{stokes0})
 and obtain
$$
\begin{pmatrix}
y_1,& y_2,& \cdots,&  y_n \\
y'_1& y'_2,& \cdots,& y'_n \\
\vdots&    & \cdots,& \vdots\\
y^{(n)}_1& y^{(n)}_2,& \cdots,& y^{(n)}_n
\end{pmatrix}
\begin{pmatrix}
\tau^{(1)}_1(\lambda) \\
\tau^{(2)}_1(\lambda) \\
\vdots\\
\tau^{(n)}_1(\lambda)
\end{pmatrix}
=
\begin{pmatrix}
y_0 -(-1)^n y_{n+1}\\
y'_0 -(-1)^n y'_{n+1} \\
\vdots\\
y^{(n)}_0-(-1)^n y^{(n)}_{n+1}
\end{pmatrix}.
$$
The application of Cramer's formula yields $\tau^{(1)}_1$ in 
the form,
$$
\tau^{(1)}_1= 
\frac{[y_2,y_3,\cdots,y_{n+1}]}{[y_1,y_2,\cdots, y_n]}
+ \frac{[y_0,y_2,\cdots, y_n]}{[y_1,y_2,\cdots,y_n]}.
$$
We use Lemma \ref{recursiondvf} to the second term in the rhs to obtain,
$$
\tau^{(1)}_1= 
\frac{[y_2,y_3,\cdots,y_{n+1}]}{[y_1,y_2,\cdots, y_n]}
+ \frac{[y_0,y_1,\cdots, y_{n-1}][y_2,y_3,\cdots, y_{n-1}]}
 {[y_1,y_2,\cdots,y_n] [y_1,y_2,\cdots,y_{n-1}]} +
 \frac{[y_0,y_2,\cdots,y_{n-1}]}{[y_1,y_2,\cdots,y_{n-1}]}.
$$
It is now obvious that repeated applications of Lemma \ref{recursiondvf}
to the last term results the expression (\ref{DVFsln}).
\end{proof}

We mimic the case of $n=1$ and introduce $D$ functions 
$$
[y_j, \cdots, y_{k+j}]|_{x=0} = q^{(n-k)(k+1)(j+k/2)/2} D^{(k+1)} (\lambda q^{j+k/2}).
$$
Then the $a+1-$th term in (\ref{DVFsln}) reads,
$$
\frac{[y_2,y_3,\cdots, y_{a+1}] [y_0,y_1,\cdots, y_{a}]}
     {[y_1,y_2,\cdots, y_{a}]   [y_1,y_2,\cdots, y_{a+1}]}
=q^{a-n/2} 
\frac{D^{(a+1)}(\lambda q^{a/2})  D^{(a)}(\lambda q^{(a+3)/2}) }
     {D^{(a+1)}(\lambda q^{(a+2)/2})  D^{(a)}(\lambda q^{(a+1)/2})}.
$$

The DVF consists of $n+1$ terms
for the solvable $U_q(A^{(1)}_1)$ model of which auxiliary space
is $\Lambda_1$ as a classical module.
It is characterized by Baxter's $Q$ operators of $n$ species,

\begin{eqnarray*}
T^{(1)}_1(\lambda)&=& 
f_{n} \frac{ Q^{(n)}(\lambda q^{(n+1)/2}) }
     { Q^{(n)}(\lambda q^{(n-1)/2})     }
+ f_{n-1} \frac{ Q^{(n)}(\lambda q^{(n-3)/2}) Q^{(n-1)}(\lambda q^{n/2})   }
       { Q^{(n)}(\lambda q^{(n-1)/2}) Q^{(n-1)}(\lambda q^{(n-2)/2})       }
+ \cdots  \\
& &+f_{a} \frac{ Q^{(a+1)}(\lambda q^{(a-2)/2}) Q^{(a)}(\lambda q^{(a+1)/2})   }
       { Q^{(a+1)}(\lambda q^{a/2}) Q^{(a)}(\lambda q^{(a-1)/2})       }
+\cdots \\
& &+ f_1 \frac{ Q^{(2)}(\lambda q^{-1/2}) Q^{(1)}(\lambda q)}
           { Q^{(2)}(\lambda q^{1/2}) Q^{(1)}(\lambda q)} 
 + f_0  \frac{ Q^{(1)}(\lambda q^{-1})}
          {Q^{(1)}(\lambda  )}     .
\end{eqnarray*}

Clearly, we have

\begin{thm}
Under the identification, $\tau^{(1)}_1(\lambda)
 \leftrightarrow T^{(1)}_1(\lambda q)$,
\begin{equation}
Q^{(a)}(\lambda ) \leftrightarrow  D^{(a)}(\lambda ) , 
\quad f_a \leftrightarrow q^{-n/2+a}
\label{identsln}
\end{equation}
two DVFs coincide. 
\end{thm}

The  pole-free property of $\tau^{(1)}_1(\lambda)$,
required from the linear independence of FSS,
results BAE,

$$
-q^{-1}=\frac{ Q^{(a-1)}(\lambda^{(a)}_j q^{-1/2})  
Q^{(a)}(\lambda^{(a)}_j q)  Q^{(a+1)}(\lambda^{(a)}_j q^{-1/2}) }
{Q^{(a-1)}(\lambda^{(a)}_j q^{1/2})  
Q^{(a)}(\lambda^{(a)}_j q^{-1})  Q^{(a+1)}(\lambda^{(a)}_j q^{1/2})
}, \qquad (a=1,2,\cdots, n)
$$
where $Q^{(a)}(\lambda_j^{(a)})=0$ and $Q^{(n+1)}=Q^{(0)}=1$.

Thus we have verified the common algebraic structure for arbitrary $n$.

The representation (\ref{DVFsln}) or 
the identification (\ref{identsln}) is , however, not unique.
One easily recognizes this by remembering the simplest case ($n=1$)
where two different expressions are available for
$\tau^{(1)}_1$.
This originates from the simple fact that both $y$ and its derivative
are solutions to Baxter's $T-Q$ relation.
The situation is also true for $n>1$.
One can show that the identification 
\begin{equation}
Q^{(a)}(\lambda q^{a+k}) \leftrightarrow  
\frac{d^{j}}{dx^j} [y_{k+1},y_{k+2},\cdots, y_{k+a}]|_{x=0}, j=0,1,\cdots
\label{identsln2}
\end{equation}
works and we have a variety of representations for the same
$\tau^{(1)}_1$.
This is shown by using formulas analogous to  Lemma \ref{recursiondvf},
and the detail will be published
elsewhere.

Before closing the section, we comment on $\tau^{(a)}_m, (a >1)$.
The corresponding DVFs are known in the integrable models, but 
 explicit forms are quite involved.
Still, one can parameterize them by analogue of
Young tableaux\cite{KSanaly}.
We  utilize 
the "tableaux" representation in proving the 
equivalence of the DVF in  integrable models
and Stokes multipliers $\tau^{(a)}_m$ for 
$a,n$ and $m$ general.
This point may be further discussed in a separate publication.

%
%
%---------------------------------------
%
%

\section {Summary and discussions}\label{sumdis}

In the present report, we discuss a curious
connection between $n+1$ the order ODE and integrable models.
When $n=1$, the connection is efficient enough to
derive analytic equations which yields estimations of 
eigenvalues and Stokes multipliers.

For higher $n$, the correspondence is still at the 
algebraic relation level.
Unfortunately, the definition of the eigenvalue problem 
is  not necessary 
clear for higher order differential equations.
The characterization of the eigenspace (it is the Hilbert space for
$n=1$) is not obvious.
More technically,
there are several subdominant solutions in each sector. 
This obscures the identification of $D(E)$ in the general Stokes
matrices.
The lack of the connection prevents us from writing down
the integral equations and evaluating parameters like "$r$"
for $n=1$.
We however comment some progress made in \cite{DT3,NewDT}

The observation made in the last few sections may be interesting.
Suppose that the ODE/IM correspondence even occurs at the construction of
models.
Then one may find the variable $x$ also in IM.
Once if one of Baxter's $Q^{(1)}$ is constructed, the other 
independent  $Q^{(1)}$ functions are 
found in derivatives of the $Q^{(1)}$ with respect to
the hidden variable $x$.
Moreover, one can generate higher $Q$ functions, i.e., $Q^{(a)}, a\ge 2$
mere by taking determinants of fundamental  $Q^{(1)}$.
The other $Q^{(a)}$'s are again obtainable via taking  derivatives.
On the other hand, construction of $Q$ functions via the standard
"pair-propagation" argument \cite{Baxbook} seems to be far more complicated 
for $n>1$.
To the authors' knowledge, the explicit construction of 
$Q$ is done only for cases corresponding $n=1$ \cite{Baxbook,KSS,Der}, 
and the procedure
is already involved.
The systematic construction found in ODE is not obvious in IM.
The present results for ODE may be a clear guide
for analyses in the analogous issue in IM,
 but it needs further research.

Finally, we comment that the ODE/IM correspondence is
still at the "phenomenological" stage.
The fundamental question as to the origin of the 
correspondence is still open.
The complete classification of the ODE
tractable with the IM approach may need the
answer to this fundamental question.

I hope to clarify these issues in future publications.

\section*{Acknowledgments}
The author thanks organizers of the conferences, 
 " Exactly Solvable Models of Statistical Mechanics
	and Mathematical Physics " (Seoul June 2000)  and
"  Development in Discrete Integrable Systems - 
	Ultra-Discretization, Quantization  "
    (Kyoto August 2000).
He is indebted to Y. Takei
for comments and discussions.
Thanks are also due to R. Tateo for drawing 
the author's attention to the preprint \cite{NewDT}
, kind correspondences and pointing out errors in 
the original manuscript, 
and to P. Dorey for comments and the explanation
on \cite{DT2, PT2}.

%\end{document}


\begin{thebibliography}{10}
%
\bibitem{DT1} P. Dorey and R. Tateo,
Anharmonic oscillators, the thermodynamics Bethe ansatz, and 
nonlinear integral equations, 
J. Phys. {\bf A 32} (1999) L419-L426,(hep-th/9812211).
%
\bibitem{DT2} P. Dorey and R. Tateo,
On the relation between Stokes multipliers and the $T-Q$ systems
of conformal field theory, 
Nucl. Phys. {\bf B 563} (1999) 573-602.
,(hep-th/9906219).
%
\bibitem{DT3}  P. Dorey and R. Tateo,
Differential equations and integrable models: the SU(3) case,
Nucl. Phys. {\bf B 571} (2000) 583-606.

%
\bibitem{BLZnew} V. V. Bazhanov, S. L. Lukyanov and A. B. Zamolodchikov,
"Spectral determinants for Schr{\"o}dinger equation and $Q-$ operators
of Conformal Field Theory", hep-th/9812247.
%
\bibitem{JS} J. Suzuki,
 Anharmonic Oscillators, Spectral Determinant and
Short Exact Sequence of $U_q(\widehat{\mathfrak{sl}}_2)$ ,
J. Phys. {\bf A 32} (1999) L183-L188,(hep-th/9902053).
%
\bibitem{SUNStokes} J. Suzuki,
 Functional Relations in Stokes multipliers and 
		 Solvable Models related to $U_q(A^{(1)}_n)$,
J. Phys. {\bf A 33} (2000) 3507-3521.
%
\bibitem{doublewell} J. Suzuki,
 Functional Relations in Stokes multipliers -fun with $x^6+\alpha x^2$
 potential, (quant-ph/0003066)
 
 
\bibitem{V1} A. Voros, 
The return of the quartic oscillator, 
 Ann. Inst. H. Poincare {\bf A39} (1983) 211-338.
%
\bibitem{V2} A. Voros,  Spectral Zeta Functions,
 Adv. Stud. Pure. Math. {\bf 21} (1992) 327-358.
%
\bibitem{V3} A. Voros,  Exact quantization condition for anharmonic
oscillators  (in one dimension), 
J. Phys. {\bf A27} (1994) 4653-4661.
%
\bibitem{V4} A. Voros, Exact anharmonic quantization condition
(in one dimension), 
  in {\sl Quasiclassical Method} (IMA Proceedings,
Minneapolis 1995) eds. J.Rauch and B. Simon, 
IMA Series {\bf 95} 189-224 (Springer 1997).
%
\bibitem{V5} A. Voros, 
Airy function (exact WKB result for potentials of odd degree),
 J. Phys. {\bf A 32} (1999) 1301-1311.
%
\bibitem{V6} A. Voros,
Exact resolution method for general 1D 
polynomial Schr{\"o}dinger equation, 
 J. Phys. {\bf A 32} (1999) 5993-6007.
%
\bibitem{V7} A. Voros, Exact quantization method for the
polynomial 1D Schr{\"o}dinger equation, 
 in  {\sl  Toward the exact WKB analysis of differential equations, 
linear or non-linear\/} eds. T. Kawai {\it et al.\/} 
(Proceedings, Kyoto 1998), to be published by Kyoto University Press.

\bibitem{V8}  A. Voros, Exercises in exact quantization,
(math-ph/0005029), J. Phys A in press.


\bibitem{Takei} T. Kawai and Y. Takei,{\sl Algebraic Analysis on Singular 
Perturbations} (Iwanami, 1999) in Japanese.


 
\bibitem{Baxbook} R. J. Baxter, {\sl Exactly Solved Models in Statistical
Mechanics} (Academic Press)

\bibitem{BLZ1} V. V. Bazhanov, S. L. Lukyanov and A. B. Zamolodchikov,
Integrable structure of conformal field theory, quantum KdV theory and
thermodynamic Bethe ansatz, 
Comm. Math. Phys. {\bf 177} (1997) 381-398.

%
\bibitem{BLZ2} V. V. Bazhanov, S. L. Lukyanov and A. B. Zamolodchikov,
Integrable structure of conformal field theory II. $Q-$ operator
and DDV equation, 
Comm. Math. Phys. {\bf 190} (1997) 247.
%
%
\bibitem{BLZ3} V. V. Bazhanov, S. L. Lukyanov and A. B. Zamolodchikov,
Integrable structure of conformal field theory III. 
The Yang Baxter relation
Comm. Math. Phys. {\bf 200} (1999) 297-324.


\bibitem{Hiro}
R. Hirota, J. Phys. Soc. Jpn. {\bf 50}(1981) 3785.

\bibitem{Miwa}

T.Miwa, Proc. Japan. Acad. {\bf 58}(1982) 9.

\bibitem{NewDT}
P. Dorey, C. Dunning and R. Tateo,
Differential equations for general $SU(n)$
Bethe ansatz systems, (hep-th/0008039).




%
\bibitem{Sbook}  Y. Sibuya, {\it Global Theory of second order linear
ordinary differential operator with a polynomial coefficient} 
Mathematics Studies 18
(North-Holland 1975).

%
\bibitem{HS} P-F Hsieh and Y. Sibuya, 
On the asymptotic integration of second order linear ordinary
differential equations with polynomial coefficients, 
J. Math. Analysis and Applications 
{\bf 16} (1966) 84-103.


		 
\bibitem{Fbook} M. V. Fedoryuk,{\it  Asymptotic analysis} (Springer 1993)


\bibitem{Reshet}
N. Reshetikhin,
The spectrum of the transfer matrices connected with
Kac-Moody algebras. Lett.Math.Phys.{\bf 14} 235-246 (1987).

\bibitem{KSanaly}  A. Kuniba and J. Suzuki, 
Comm. Math. Phys.173 (1995) 225.


\bibitem{KP} A. Kl{\"u}mper and P. A. Pearce,
 Conformal weights of RSOS lattice models and 
their fusion hierarchies, 
Physica {\bf A183} (1992) 304-350.


\bibitem{JKSfusion}
G.~J{\"u}ttner, A.~Kl{\"u}mper and J.~Suzuki, 
 From fusion hierarchy to excited state TBA,  
 Nucl Phys {\bf B512} (1998) 581-600.
 
 \bibitem{PT1}
C.M. Bender and S. Boettcher,
Real Spectra in Non-Hermitian Hamiltonians
Having ${\cal P}{\cal T}$ symmetry, 
Phys. Rev. Lett.{\bf 80} (1998) 5243 .


\bibitem{PT2}
C.M. Bender and S. Boettcher and P. N. Meisinger,
${\cal PT}$-Symmetric Quantum Mechanics,
J. Math. Phys .{\bf 40} (1999) 2201 .


\bibitem{Hirota}
R. Hirota, Mathematics of Soliton by the direct method,
(in Japanese)
(Iwanami 1992)

\bibitem{KSS}
V.B. Kuznetov, M. Salerno and E.K. Sklyanin,
Quantum Backlund transformation for the integrable mDST model,
J. Phys. {\bf A33} (2000) 171-189.

\bibitem{Der}
S. E. Derkachov, Baxter's $Q-$ operator for the homogeneous XXX
spin chain, J. Phys. {\bf A32} (2000) 5299-5316.


\end{thebibliography}
\end{document}